# Materials processing with intense pulsed ion beams and masked targets

John J. Barnard[1,2] and Thomas Schenkel[2]
[1]Lawrence Livermore National Laboratory, Livermore, CA
[2]Lawrence Berkeley National Laboratory, Berkeley, CA

**Abstract**
Intense, pulsed ion beams locally heat materials and deliver dense electronic excitations that can induce materials modifications and phase transitions. Materials properties can potentially be stabilized by rapid quenching. Pulsed ion beams with (sub-) ns pulse lengths have recently become available for materials processing. Here, we optimize mask geometries for local modification of materials by intense ion pulses. The goal is to rapidly excite targets volumetrically to the point where a phase transition or local lattice reconstruction is induced followed by rapid cooling that stabilizes desired materials properties fast enough before the target is altered or damaged by e. g. hydrodynamic expansion. We performed HYDRA simulations that calculate peak temperatures for a series of excitation conditions and cooling rates of silicon targets with micro-structured masks and compare these to a simple analytical model. The model gives scaling laws that can guide the design of targets over a wide range of pulsed ion beam parameters.

## I. Introduction

Beams of energetic ions have been used to alter materials properties for many years, with doping of semiconductors by ion implantation as one leading application [1, 2]. Dynamic annealing processes induced by energetic ions can aid damage repair [3, 4] but targets, such as semiconductor wafers, are generally annealed thermally following ion implantation in order to repair lattice damage and to ensure substitutional incorporation of dopants in the host crystal matrix. In the early 1980s short intense ion pulses with pulse lengths of 50 to 100 ns became available where the pulse energy was high enough, 1 to 10 J/cm$^2$, to simultaneously implant and anneal materials [5]. Ion beam modification of materials with intense pulsed ion beams continues to be applied in areas where the characteristics of ion-solid interactions provide complementary advantages vs. other sources of directed energy for materials processing (such as laser or electron beams) [2, 6]. More recently, intense ion beams with much shorter pulse lengths of tens of ps to a few ns have become available for materials studies. These ion beams are based on laser-plasma acceleration of ions [7] or based on induction accelerator technology [8, 9]. The much shorter pulse lengths can now enable access to new materials processing regimes and we explore this with simulations of the thermal response of materials to intense, short ion pulses. In masked targets, the heating and cooling rate can be tuned by the mask geometry together with the ion pulse parameters. We investigate process parameters for intense ion pulses for applications where the characteristics of ion-solid interaction are advantageous vs. complementary processing schemes with (ultra-) short pulse lasers [2, 7], where a key differences for ion vs. laser pulses is the uniform energy deposition that can be achieved with energetic ions over many microns for any material.

We view intense ion pulses with masked targets as a processing opportunity that combines aspects of two more common cases. One is the interaction of single, swift heavy ions with materials, where local phase transitions and materials modifications can be induced by the very high rate of electronic energy loss (e. g. 20 keV/nm for 1 GeV gold ions in silicon). Local heating and electronic excitation on a nanometer scale can drive materials well above the melting point for a few picoseconds followed by very rapid cooling and quenching of the intense, ultrafast electronic excitation [10, 11], driving complex materials dynamics such as track formation and defect annealing [12, 13]. When combined with external pressure, novel phases can be induced and stabilized [14] and color centers, such as nitrogen vacancy centers in diamond can be formed locally and without need for subsequent thermal annealing [15]. The other common scenario is to simply use a broad beam of intense ions [6]. Now heat conduction is mostly limited to the depth of the target and materials simply evaporate or explode due to hydrodynamic expansion on a time scale roughly given by the ratio of the ion range and the speed of sound in the material, e. g. a few ns for excitation of solid targets over a depth of a few microns [16, 17]. Here, we show simulations and an analytical model that inform mask design for local heating of materials by intense ion pulses with optimized heating and cooling rates. We illustrate the three regimes of pulsed ion beam processing of materials in Figure 1.

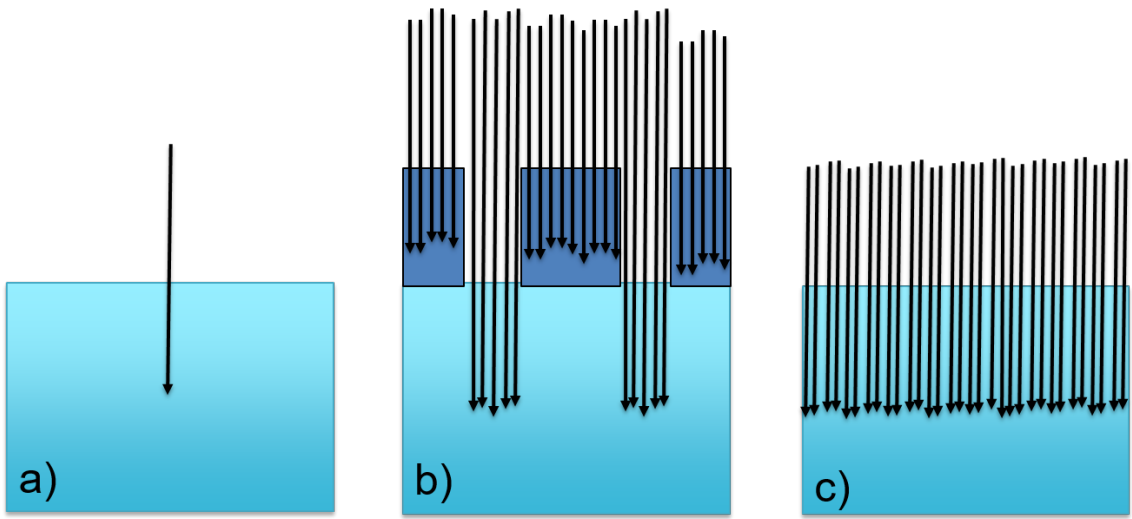

**Figure 1.** Illustration of processing regimes. a) a single, swift heavy ion (nm and ps scale heating and cooling), b) intense, short (ps to ns) ion pulse with (sub-) micron masks to optimize heating and cooling rates, and c) ion beam pulse without masking, where cooling is limited to one dimension.

Our article is motivated by the intriguing prospects of extreme chemistry and phase engineering where one could drive materials such as silicon [19], diamond [10, 15] or complex oxides [14] to extreme conditions of electronic excitation density, temperature and pressure and then stabilize novel materials properties by rapid quenching. By using a mask, the longitudinal dimension can be large compared to the transverse dimension, allowing the possibility of rapid transverse cooling. We explore the tradeoffs between mask hole size, beam pulse duration, and beam intensity such that target temperature and cooling rate (after heating) are both maximized.

**II. Model**

Figure 2 illustrates the basic geometrical configuration that we are considering. For concreteness, we use 1.2 MeV helium ion beam pulses (as routinely available at NDCX-II, an induction accelerator a Berkeley Lab [8, 9]), a gold mask, and a silicon target. The ion beam illuminates the high density, metallic mask that has small holes (0.1 to 1 µm radius) which allow the ion beam to pass and reach the target. The mask is sufficiently thick (~ 3 µm of gold) so that it can absorb the beam pulse energy and so that the ions are stopped in the mask before reaching the target. The beam then heats a nearly uniform cylinder of target material that has a length that is given by the ion energy deposition profile, (ca. 4 µm for 1.2 MeV He+ in silicon) and that is much longer than the radius that is defined by the mask. We neglect ion straggling here, which will broaden this radius slightly.

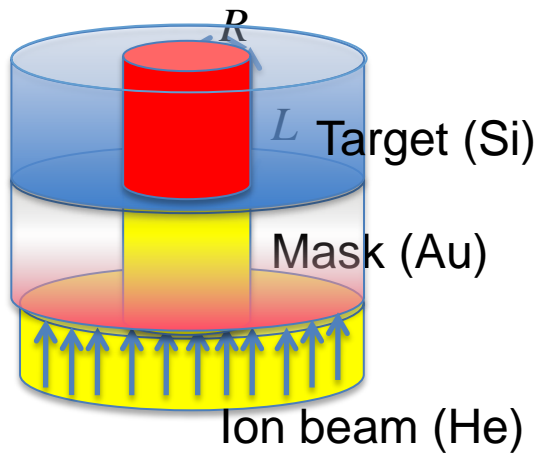

**Figure 2.** Geometry of ion beam, mask, and target. The (He+) ion beam (full beam radius ~1 mm) impinges upon a (gold) mask absorbing most of the beam energy, but allowing a ~0.1 µ to ~1 µ radius ion beam to pass to the silicon target. For a 1.2 MeV He ion beam the gold mask must be ca. 2 to 5 microns thick.

We note that geometries for the mask that we consider do not have to be restricted to single beam openings. Multiple channels can be used as well as long as the spacing between channels is sufficient so that cooling from one channel is not affected by adjacent channels, thus spacings should be larger than the hole diameters.  Our analysis can also be extended to arbitrary mask geometries.

### III. Simulation Result

We have carried out simulations using the hydrodynamics code HYDRA [18].  In addition to the Au mask (with 0.2 μ hole) and Si target, the hole and region facing the mask was filled with a low density argon gas for simulation convenience. The gas was low enough density (density ~ $10^{-9}$ g/cm$^3$) so that it did not impact the dynamics. The ion beam had a parabolic temporal profile with full width of 0.8 ns, and total integrated fluence of 0.4 J/cm². Figure 3 a) shows the target density profiles at the end of an 0.8 ns ion pulse where the gold mask material had expanded slightly into the hole volume, reducing the fluence delivered to the un-masked silicon target area by about 20%.  Figure 3 b) shows a snapshot of the corresponding temperature profile 0.5 ns into the ion pulse.  Here, we see that the gold foil prevented heating of the silicon target as expected at large radius and that a warm region (~0.036 eV, or 690 K) was created in the silicon volume that had not been covered by the gold mask.. Figure 4 summarizes the time history of the un-masked, central silicon temperature for four different mask hole radii of 0.1, 0.2, 0.4, and 1.0 μm.

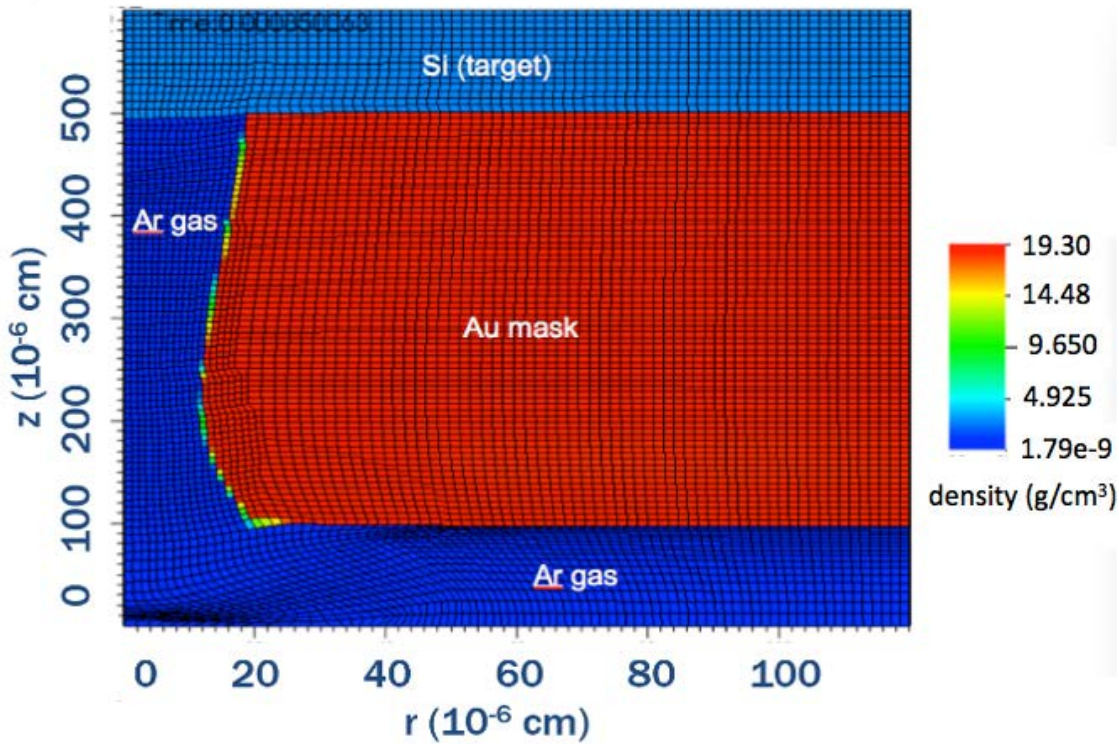

(a)

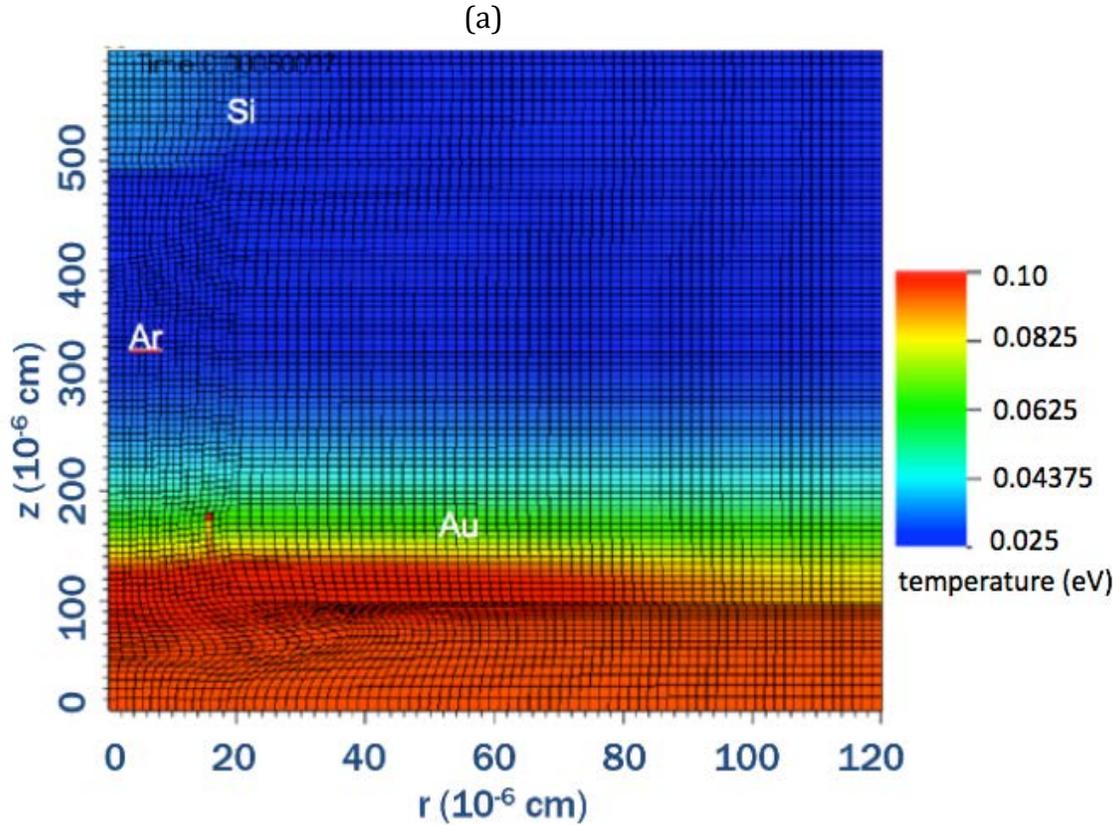

(b)
**Figure 3**. (a) False color map of density of a silicon target (light blue[top]), gold mask (red), with 0.2 micron hole filled with very low density Ar gas heated by 1.2 MeV He ion beam with parabolic intensity profile in time, illuminating the target from below at the end of the heating pulse that lasted 0.8 ns. (b) temperature profiles (in eV) near peak temperature at 0.5 ns.

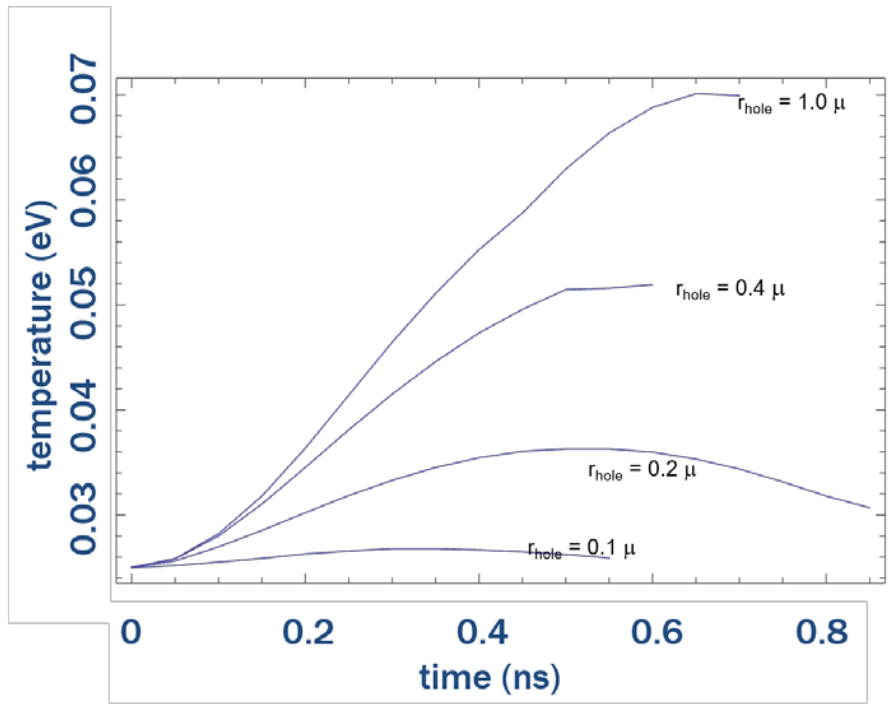

**Figure 4.** Evolution of central temperature in the silicon target vs. time for four different hole radii. For all four curves, the ion beam (1.2 MeV He⁺) had a pulse duration of 0.8 ns and a fluence on target of 0.4 J/cm².

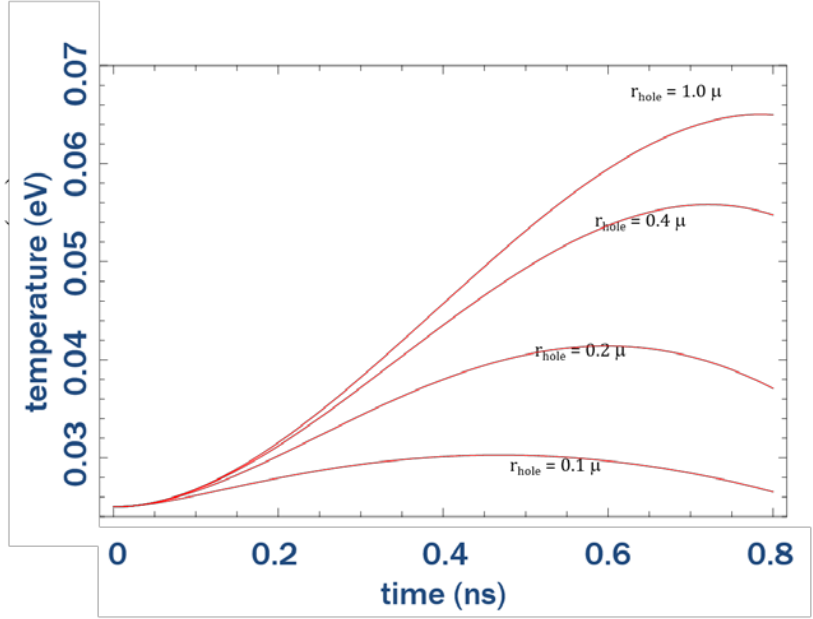

**Figure 5.** Application of simple analytic model, discussed in section IV to parameters of figure 4. Differences between results in the two figures are due to hole closing not being considered in the analytical model, ; the ion range, L, was now estimated from SRIM [20] rather than HYDRA and a simplified constant specific heat and conductivity model were used.

## IV. Analytic Model Results

In order to quickly understand the scaling of the achievable target temperatures and cooling rates obtainable from these mask geometries, we derive a simplified model that incorporates the principal physics of these masked targets. Since the heated portion of the target is primarily cylindrical (with radius $R$ and length $L$), with L > R, we estimate the evolution of the temperature, by equating the rate of change of the internal energy by the heating rate from the ion beam, less the cooling rate from conductive losses through the boundaries.

$$\pi R^2 L \frac{d}{dt}\left(\frac{3\rho k_B T}{m}\right) = \text{Heating} - \text{Cooling}$$

$$= \pi R^2 L \frac{F}{L\Delta t} \frac{3}{2}\left(1 - \frac{(t - \Delta t/2)^2}{\Delta t^2/4}\right) - (2\pi RL)\kappa \nabla T \qquad (1)$$

$$\approx \pi R^2 L \frac{F}{L\Delta t} \frac{3}{2}\left(1 - \frac{(t - \Delta t/2)^2}{\Delta t^2/4}\right) - (2\pi RL)\frac{\kappa T}{R}$$

Here $R$ = radius of mask hole, $L$ = ion stopping length, $\Delta t$ = pulse duration of the ion beam (full width of parabolic pulse), $k_BT$ = target (silicon or diamond) temperature (in eV, with $k_B$ Boltzmann constant), $\rho, m$ = target (silicon or diamond) density, mass, $\kappa$ = target thermal conductivity, and $F$ = incident ion fluence (energy/area). Note that if $R$ is large then the left hand side in eq 1 is balanced by heating and the maximum target temperature occurs at the end of the beam pulse when $t= \Delta t$:

$$k_B T_{max 1} = \frac{Fm}{3\rho L} \quad \text{at } t = \Delta t \quad \text{assuming } R > \frac{2}{3}\left(\frac{\kappa m \Delta t}{k_B \rho}\right)^{1/2} \equiv R_c \quad (2).$$

However, when R is small, heating is always just balanced by cooling and the temperature follows the instantaneous beam intensity during the ion beam pulse, which for the case of a parabolic pulse occurs at t $t= \Delta t/2$:

$$k_B T_{max 2} = \frac{3 k_B F R^2}{4 \kappa L \Delta t} \quad \text{at } t = \Delta t/2 \text{ assuming } R < \frac{2}{3}\left(\frac{\kappa m \Delta t}{k_B \rho}\right)^{1/2} \equiv R_c \quad (3).$$

The simple model has the form:

$$\frac{d k_B T}{dt} + \frac{k_B T}{\tau} = \frac{k_B T_{max 1}}{\Delta t}\left[\frac{3}{2}\left(1 - \frac{(t - \Delta t/2)^2}{\Delta t^2 /4}\right)\right],$$

and can be solved exactly, with solutions during and after the ion pulse:

$$\frac{\Delta k_B T(t)}{k_B T_{max 1}} = \begin{cases} 6\left(\frac{\tau^2}{\Delta t^2}\right)\left[\left(1+\frac{2\tau}{\Delta t}\right)(t/\tau + \exp(-t/\tau) - 1) - \frac{t^2}{\tau \Delta t}\right] & \text{for } 0 < t < \Delta t \\ 6\left(\frac{\tau^2}{\Delta t^2}\right)\left[\left(1+\frac{2\tau}{\Delta t}\right)(\Delta t/\tau + \exp(-\Delta t/\tau) - 1) - \Delta t/\tau\right]\exp(-t/\tau) & \text{for } t > \Delta t \end{cases}$$

(4).

Here $\tau = \frac{3 k_B \rho R^2}{2 \kappa m} = \frac{4}{9}\frac{k_B T_{max 2}}{k_B T_{max 1}}\Delta t$ is the cooling time, and from the definitions $\frac{R}{R_c} = \sqrt{\frac{3\tau}{2\Delta t}}$, where we introduce the ratio of a given mask radius, R, to a critical mask radius, $R_c$, to highlight the connection of mask geometry to the time structure, given by the ration of the cooling time, $\tau$, and the ion pulse duration, $\Delta t$.

In eq. (4) the normalized temperature, $T(t)/T_{max1}$, depends only on time relative to the pulse duration, $t/\Delta t$, and on the ratio of the cooling time to the pulse duration, $\tau/\Delta t$. In order to visualize this scaling, we now plot the normalized temperature and its normalized time derivative in figures 6 and 7. We also plot the time of maximum temperature and the values of maximum temperatures in figures 8a and 8b as a function of $\tau/\Delta t$ (the ratio of the cooling time to the pule duration). We further show the maximum normalized time derivative of the normalized temperature (also as a function of $\tau/\Delta t$ ) in figure 9.

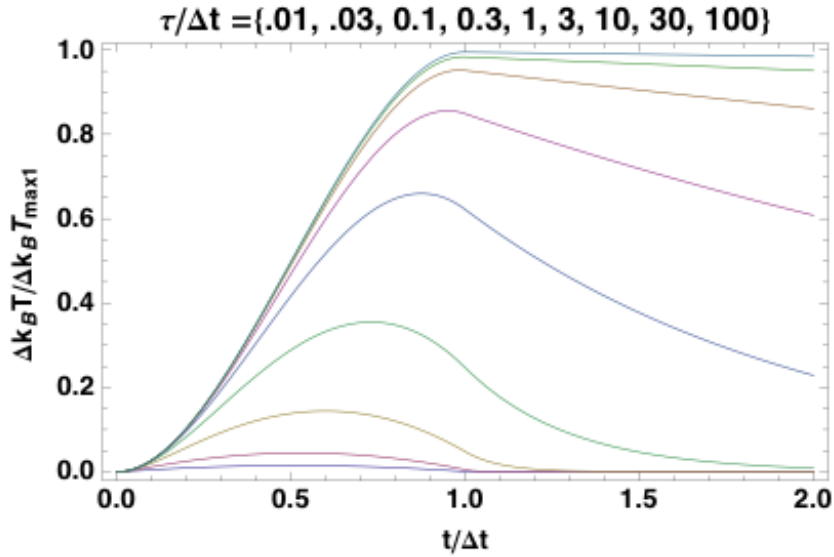

**Figure 6.** Evolution of normalized target temperature ($\Delta k_B T / k_B T_{max1}$) for nine values of $\tau/\Delta t$.

Figure 6 shows that, as expected, for very low cooling rates, cooling times, $\tau$, are very long compared to the pulse length, $\Delta t$, and a high steady state temperature is reached for $\tau/\Delta t > 10$. At the other extreme heating is always balanced by very high cooling rates and short cooling times, $\tau/\Delta t < 0.3$, and the temperature hardly rises. The most interesting case is intermediary, where heating and cooling are balanced during an ion pulse and a distinct peak temperature is reached for $\tau/\Delta t \sim 1$.

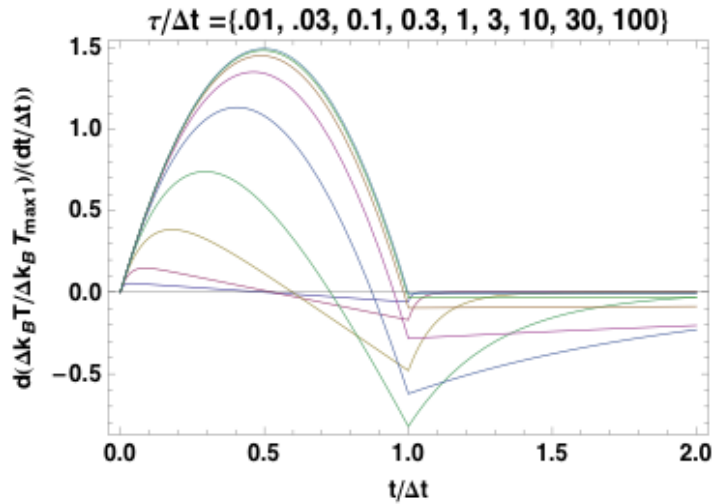

**Figure 7.** Evolution of derivative of normalized target temperature $d(\Delta k_B T / k_B T_{max1})/d(t/\Delta t)$ for nine values of $\tau/\Delta t$, the ration of the cooling time to the pulse duration. Note that the maximum cooling rate occurs for all curves at $t/\Delta t = 1$ (at the end of the heating pulse).

In figure 7 we show the time derivative of the relative temperature change from Figure 6, thus the rate of temperature change as a function of time during the ion pulse, t/Δt. Here, positive values of the derivative indicate heating and negative values show net cooling. This derivative is useful in highlighting a maximum cooling rate for a ratio of cooling time to pulse duration of $\tau/\Delta t=0.3$.

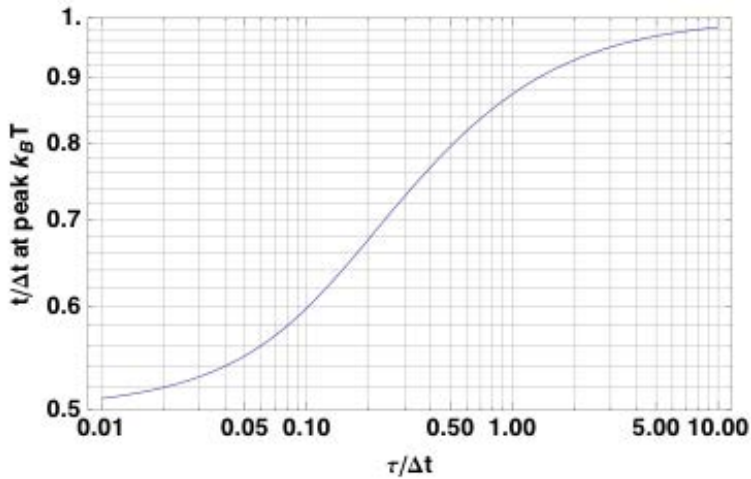

(a)

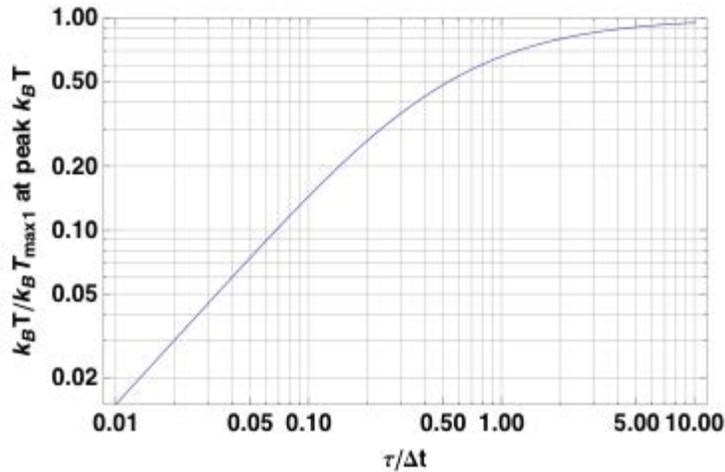

(b)

**Figure 8.** (a) Time of maximum normalized temperature and (b) value of maximum normalized temperature as a function of $\tau/\Delta t$. Note that at small $\tau/\Delta t$ (small hole radii R) the heating balances cooling, so that the temperature follows the ion beam intensity, which is maximum at $t/\Delta t =1/2$, whereas at large $\tau/\Delta t$ (large hole radii R) cooling is negligible and the maximum temperature just depends on the accumulated fluence, and so the temperature is maximum at the end of the pulse $t/\Delta t =1$. In the latter case the maximum temperature is just $T_{max1}$.

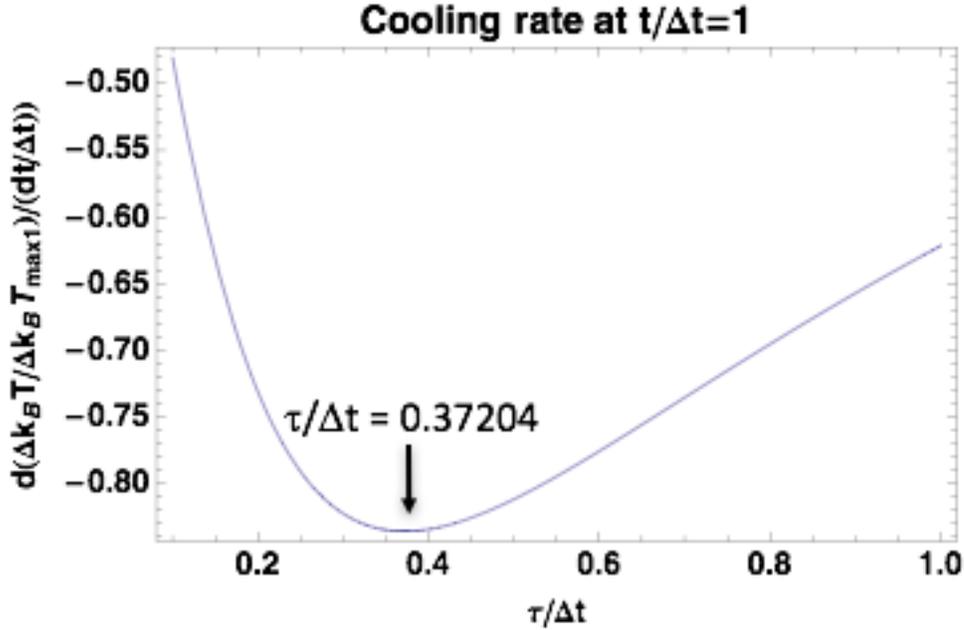

**Figure 9.** Maximum normalized time derivative (at $t/\Delta t =1$) of the normalized temperature as function as a function of $\tau/\Delta t$. The maximum cooling rate occurs at $\tau/\Delta t = 0.37$, and the maximum normalized temperature $\Delta k_B T/k_B T_{max1} = 0.41$ for this value of $\tau/\Delta t$, at a time $t/\Delta t= 0.76$.

## V. Discussion

Implementation of rapid local heating followed by rapid quenching of a material using masked targets requires a choice of hole radius and pulse duration that maximizes both the temperature achieved during the ion pulse and the cooling rate after the ions have been delivered.
As an example in figures 10 and 11 we use the simple model (eq. 4) to explore these tradeoffs for silicon targets. In these two figures, we fix the desired target temperature at 0.5 eV (about 5800 K). We then calculate what fluence is required (figure 10) as a function of hole radius, for three pulse durations, for the combination of a Si target ($\kappa$ = 1.49 W/(cm deg C, $\rho$ = 2.3 g/cm$^3$) and a 1.2 MeV He ion beam (L = 4.2 μm). We further determine the cooling rate that is achieved at the needed fluence as a function of hole radius (figure 11). We also note that HYDRA simulations predict that pressure increases to about 7 GPa are associated with lattice temperatures rises to 0.5 eV driven by these ion pulse conditions [17].

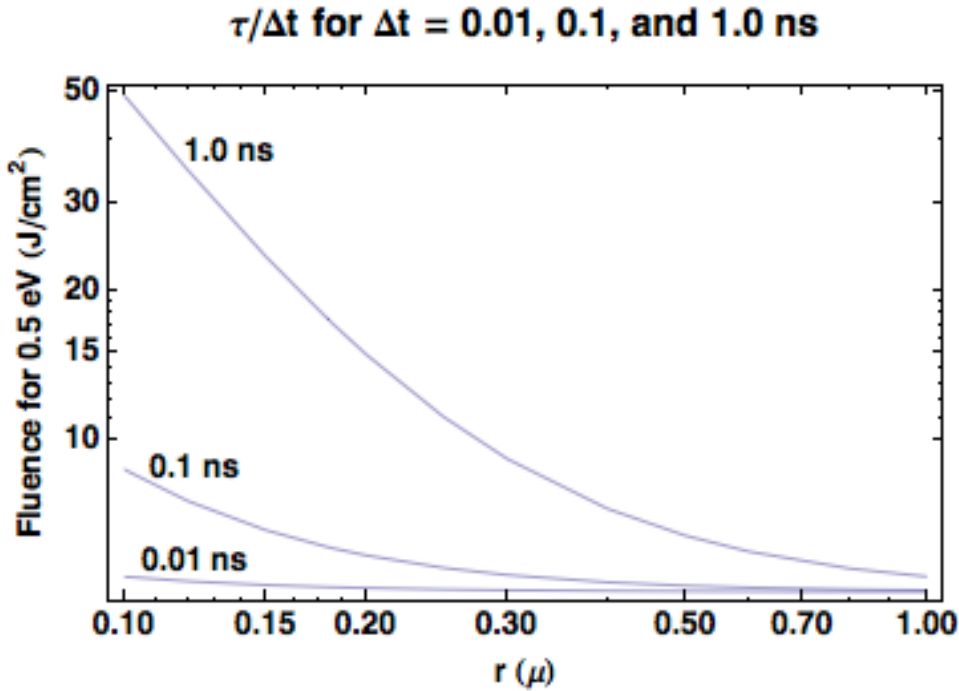

**Figure 10.** Required fluence to reach a maximum target temperature of 0.5 eV as a function of hole radius, for three ion pulse durations using the model of equation (4) and a Silicon target illuminated by a 1.2 MeV ion beam.

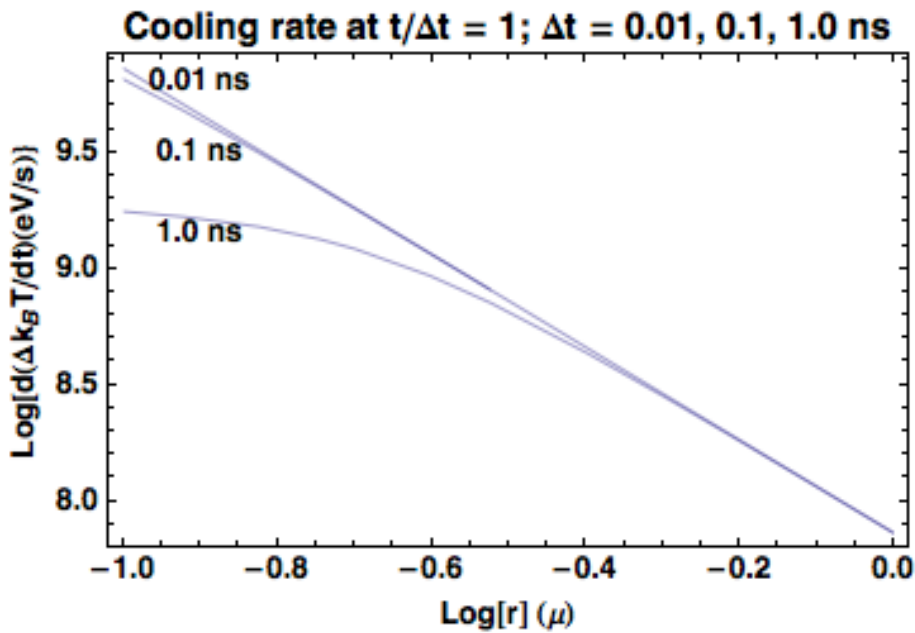

**Figure 11.** Cooling rate at the end of the beam pulse for a target reaching a maximum temperature of 0.5 eV as a function of hole radius, for three ion pulse durations using the model of equation (4) and a Silicon target illuminated by a 1.2 MeV ion beam.

As can be seen from figures 10 and 11 (and 9) if R << $R_c$ (i.e. the cooling time is much shorter than the ion pulse length, τ/Δt << 1), then the fluence required to reach the desired temperature goal is much larger than in the opposite extreme when R >> $R_c$ (i. e. the cooling time is much longer than the pulse length, τ/Δt >>1). However, as R/$R_c$ increases the cooling rate decreases. It is likely that the desirable operating point would be in the range where τ/Δt ~ 0.4 -1 near the knee of the curve in Figure 9 where cooling rates are large due to relatively small hole radii and where there is not too much of a penalty yet in the required fluence. Importantly, we also see that shorter ion pulses optimize both maximum temperature and maximum post-heating cooling rate.

We may now compare heating and beam parameters for two example ion beams: an accelerator produced He$^+$ ion pulse from NDCX-II (the neutralized drift compression experiment at Berkley Lab) [8, 9] and a laser generated proton pulse from a petawatt laser such as Bella-i at Berkeley Lab [7, 21]. NDCX-II can produce ion energy fluences of 0.4 to 0.8 J/cm$^2$ in pulse durations of order 2 ns, and we expect Bella-i will be able to produce much higher fluences of order 100 J/cm$^2$ on similar time scales of a few ns after ballistic de-bunching of the ion pulse with broad energy distribution. Ion pulses are initially not much longer than the driving laser pulse [22], but ion pulse de-bunch quickly due to large ion energy distributions and sub-ns pulse durations can be restored by re-bunching [23]. Because the ion energies are higher in Bella-i and ion ranges increase in proportion to the increased ion energy the expected temperatures are only modestly higher, but the decreased pulse duration (with compression) will allow an order of magnitude higher cooling rate, and exploration of lattice relaxation on the 10 to 100 ps time scale. Considerably higher lattice temperatures than for proton pulses are obtainable using higher mass ions such as carbon. With our two approaches to ion pulse generation and this target geometry, lattice relaxation can be studied (both in situ and ex-situ) with both heating *and* cooling times of order the pulse duration (between 10 ps and a few ns).

Ion beam driven heating of a target lattice results from the deposition of kinetic energy through relatively well known elastic and inelastic scattering processes. For 1.2 MeV He ions, electronic stopping processes account for about 99% of energy loss processes. Energetic electrons cascade and couple to the atomic lattice through elastic and inelastic collisions and effective electron – phonon coupling dynamics, resulting in target lattice heating [11]. We have explored how we can control local heating through the balance of heating and cooling with a simple mask geometry for short, intense ion pulses. We have found that this is indeed promising because the characteristic length scale for cooling during (sub)-ns – scale pulses is of order 0.1 to 1 micron, i. e. in a range readily accessible by standard lithography techniques. The range of delta electrons (energetic secondary electrons from inelastic scattering) can be much shorter than mask openings. For 1.2 MeV He$^+$, the cut-off energy for delta electrons is about 600 eV with a range of only about 10 nm in silicon [25].  The rate of local electronic excitation is thus not affected by masking.  For 10 MeV protons, the cut-off delta-electron energy is about 20 keV with a range of about 1 micron in silicon, now comparable to or larger than possible mask dimensions.   For high energy protons with ranges in excess of 50 micron in e. g. gold layers, fabrication of high aspect ratio mask structures with ratios of mask thickness to mask opening >50 also becomes challenging.  Mask aspect ratio limits favor a few MeV protons or helium ions.  Clearly, the balance of beam induced damage to the tailoring of desired properties will have to be optimized for any ion pulse condition.   Further, mixed excitation by broad proton beams (which penetrate a mask) combined with masked heavy ion beams can be considered.

We note that the peak temperature of 690 K in the example of Figure 2 is well below the temperature for thermal re-crystallization of silicon of about 850 K [1].  However, light ion beam enhanced annealing effects [3] can enhance damage repair and re-crystallization at much lower sample temperatures.  Masked exposures of materials with intense, pulsed ion beams now enable us to control the rate of local electronic excitations partially decoupled from beam induced lattice heating.  This is exciting, because it can enable e. g. local damage repair and activation of dopants that had been implanted in a prior processing step with minimal thermal budget and thus minimal diffusion or segregation to a nearby interface.  Further, repeated and iterative steps of local processing such as addition of vacancies and local excitation can enable the formation of ordered arrays of dopants and color centers, where current formation efficiencies are too low for the formation of networks of coupled spins [25, 26].

## VI. Conclusion

We have used HYDRA to simulate a silicon target heated by ion beam deposition through a gold mask of various hole sizes, and with ion beams of various pulse durations and beam fluences.  The simulations and a simple heating/cooling model show that when the hole radius is significantly greater than a critical mask radius, $R_c$, (corresponding to the cooling time $\tau$ being significantly greater than the pulse duration $\Delta t$) the central temperature is close to $T_{max1}$ (the temperature determined by the beam fluence alone, without any masking). To maximize the cooling rate, the radius can be chosen such that $\tau \sim 0.37\ \Delta t$ (corresponding to $R \sim 0.75\ R_c$) and this yields a maximum temperature $k_B T_{max} \sim 0.41\ k_B T_{max1}$. Intermediate choices such that $\tau \sim \Delta t$ (where the cooling time equals the pulse duration) may be made, if minimizing fluence is weighed more heavily than maximizing cooling rate. For very small radii, hole closing reduces the effective fluence on the target and must be included in the calculation. This work is of interest when rapid cooling is desired to quench and stabilize a change in phase or lattice configuration of a material that was induced by intense ion beam pulses.

**Acknowledgments:**

This work was supported by the US DOE under contracts DE-AC02-05CH11231 (LBNL) and DE-AC52- 07NA27344 (LLNL).